\documentstyle[12pt]{article}

\begin{document}

\author{A. de Souza Dutra\thanks{%
E-mail: dutra@feg.unesp.br} \\
$^{a}$Abdus Salam ICTP, Strada Costiera 11, 34014 Trieste Italy\\
$^{b}$UNESP-Campus de Guaratinguet\'{a}-DFQ\thanks{%
Permanent Institution}\\
Av. Dr. Ariberto Pereira da Cunha, 333\\
C.P. 205\\
12516-410 Guaratinguet\'{a} SP Brasil}
\title{{\LARGE Quantum propagator for some classes of three-dimensional three-body
systems}}
\maketitle

\begin{abstract}
In this work we solve exactly a class of three-body propagators for the most
general quadratic interactions in the coordinates, for arbitrary masses and
couplings. This is done both for the constant as the time-dependent
couplings and masses, by using the Feynman path integral formalism. Finally
the energy spectrum and the eigenfunctions are recovered from the
propagators.


\end{abstract}

\newpage

\section{Introduction}

In a very recent work, A. Chouchaoui \cite{3bodes} calculated the propagator
for the problem of three identical particles in one dimension. This was done
by working in the Feynman path integral formalism \cite{feynman}\cite
{grosche}, so expanding the class of three-body systems solved through this
formalism as, for instance, those discussed by Khandekhar and collaborators
\cite{khandekar}, Govaerts \cite{govaerts}, and others.

On the other hand, the interest in solving problems involving time-dependent
systems has attracted the attention of physicists since a long time. This
happens due to its applicability for the understanding of many problems in
quantum optics, quantum chemistry and others areas of physics \cite{abdalla}-%
\cite{ioan}. In particular we can cite the case of the electromagnetic field
intensities in a Fabry-P\'{e}rot cavity \cite{abdalla}. In fact this kind of
problem still represents a line of investigation which attract the interest
of physicists \cite{feng}-\cite{epl05}.

Here we intend to expand the class of exactly solvable path integral
problems, by including the case of a general three-body quadratic
interaction in the coodinates, both in the case of constant as in the case
of time-dependent couplings. This is going to be done through a suitable
combination of the Jacobi coordinates with a further decoupling one.

\section{A general exactly solvable quadratic interactions for three-body
with constant couplings}

The model which we are going to treat is represented by the following
Lagrangian
\begin{eqnarray}
{\cal L} &=&\sum_{j=1}^{3}\frac{m_{j}}{2}\left( \frac{\,d\vec{r}_{j}}{dt}%
\right) ^{2}-\frac{1}{2}\left[ K_{21}\,\left( \vec{r}_{2}-\vec{r}_{1}\right)
^{2}+K_{31}\,\left( \vec{r}_{3}-\vec{r}_{1}\right) ^{2}+K_{32}\,\left( \vec{r%
}_{3}-\vec{r}_{2}\right) ^{2}+\right.   \nonumber \\
&&+\,\sigma _{1}\left( \vec{r}_{2}-\vec{r}_{1}\right) \cdot \left( \vec{r}%
_{3}-\vec{r}_{1}\right) +\sigma _{2}\left( \vec{r}_{2}-\vec{r}_{1}\right)
\cdot \left( \vec{r}_{3}-\vec{r}_{2}\right) +  \label{l1} \\
&&\left. +\sigma _{3}\left( \vec{r}_{3}-\vec{r}_{1}\right) \cdot \left( \vec{%
r}_{3}-\vec{r}_{2}\right) +\vec{g}_{1}\cdot \left( \vec{r}_{2}-\vec{r}%
_{1}\right) +\vec{g}_{2}\cdot \left( \vec{r}_{3}-\vec{r}_{1}\right) +\vec{g}%
_{3}\cdot \left( \vec{r}_{3}-\vec{r}_{2}\right) \,\right] .  \nonumber
\end{eqnarray}

As far as we know, this problem was considered only for particular cases, in
general without the crossed terms proportional to the $\sigma $ coupling
constants \cite{castro3body}, and most part of time for one-dimensional
identical particles under isotropic harmonic interactions \cite{3bodes},\cite
{khandekar}, \cite{govaerts}. As a matter of fact, we should say that, in
fact the parameters $\vec{g}_{1},\;\vec{g}_{2}$ and $\vec{g}_{3}$ can really
be time-dependent as we are going to see below in the text.

For the first step, in order to reach our goal, we perform a change of
coordinates system going to the so called Jacobi coordinates, which are
characterized in this case by the following set of transformations \cite
{castro3body}:
\begin{equation}
\left(
\begin{array}{l}
\vec{X}_{1} \\
\vec{X}_{2} \\
\vec{X}_{3}
\end{array}
\right) =\,\left(
\begin{array}{lll}
-a & a & 0 \\
- \frac{b\, m_{1}}{m_{12}} & - \frac{b\, m_{2}}{m_{12}} & b \\
\frac{m_{1}}{M} & \frac{m_{2}}{M} & \frac{m_{3}}{M}
\end{array}
\right) \left(
\begin{array}{l}
\vec{r}_{1} \\
\vec{r}_{2} \\
\vec{r}_{3}
\end{array}
\right) ,  \label{t1}
\end{equation}

\noindent where $a$, $b$ are arbitrary constants, used in order to put the
Lagrangian into a more convenient form, and $M\equiv m_{1}+m_{2}+m_{3}$ is
the total mass of the system. Note that for the sake of path-integral
evaluation, we must compute the Jacobian of the transformation due to the
impact of the transformation over the measure of the path-integral, and in
for this transformation it is given simply by $\left( ab\right) ^{3}$. As
observed in \cite{castro3body}, this transformation is not sufficient to
decouple the system, which will looks like

\begin{eqnarray}
{\cal L} &=&\sum_{j=1}^{3}\frac{M_{j}}{2}\left( \frac{\,d\vec{X}_{j}}{dt}%
\right) ^{2}-\frac{1}{2}\left[ M_{1}\omega _{1}^{2}\,\bar{X}%
_{1}^{2}+M_{2}\omega _{2}^{2}\,\vec{X}_{2}^{2}+\lambda \,\,\,\vec{X}%
_{1}\cdot \vec{X}_{2}+\right.  \nonumber \\
&&  \label{l2} \\
&&\left. +\,\vec{f}_{1}\left( t\right) \cdot \vec{X}_{1}+\vec{f}_{2}\left(
t\right) \cdot \vec{X}_{2}\right] ,  \nonumber
\end{eqnarray}

\noindent with the resulting masses and coupling constants given by
\begin{eqnarray}
M_{1} &\equiv &\frac{m_{1}m_{2}}{a^{2}m_{12}};\,\,M_{2}\equiv \frac{%
m_{3}\,m_{12}}{b^{2}\,M};\,M_{3}\equiv M;  \nonumber \\
\omega _{1}^{2} &\equiv &\frac{m_{12}}{m_{1}m_{2}}\left\{ K_{21}+\frac{1}{%
m_{12}^{2}}\left( m_{2}^{2}K_{31}+m_{1}^{2}K_{32}\right) +\right.  \nonumber
\\
&&\left. +\frac{2}{m_{12}}\left[ \sigma _{1}m_{2}-\sigma _{2}m_{1}-\sigma
_{3}\left( \frac{m_{1}m_{2}}{m_{12}}\right) \right] \right\} ; \\
\omega _{2}^{2} &\equiv &\frac{M}{m_{3}m_{12}}\left( K_{31}+K_{32}+2\,\sigma
_{3}\right) ;  \nonumber \\
\lambda &\equiv &\frac{1}{a\,b}\left[ \frac{\sigma _{3}}{m_{12}}\left(
m_{2}-m_{1}\right) +\sigma _{1}+\sigma _{2}+\frac{1}{m_{12}}\left(
m_{2}K_{31}-m_{1}K_{32}\right) \right] ;  \nonumber \\
\vec{f}_{1} &\equiv &\frac{1}{a}\left( {\large \vec{g}}_{1}+\frac{\left(
m_{2}{\large \vec{g}}_{2}-m_{1}{\large \vec{g}}_{3}\right) }{m_{12}}\right)
;\,\vec{f}_{2}\equiv \frac{1}{b}\left( {\large \vec{g}}_{2}+{\large \vec{g}}%
_{3}\right) ;  \nonumber
\end{eqnarray}

\noindent with $m_{12}\equiv m_{1}+m_{2}$.

Note that the center of mass coordinate has decoupled entirely from the
other ones, so that one can now worry only about the remaining coordinate
variables. Now one can take two possible routes. The first, which is the one
usually taken, sometimes with errors as observed in \cite{castro3body}, by
particularizing the problem through a constraint over the parameters: $%
\lambda \equiv 0$. Another possible and more general route, is the one we
are going to choose here. In such route we perform a further transformation,
as proposed in \cite{dutratransf} and used also in \cite{castro3body},
keeping in mind the need of decoupling the final system through the
elimination of the crossed term. This is reached by a simple additional
dilation and rotation transformation \cite{dutratransf}, which in the
present situation is defined as
\begin{equation}
\left(
\begin{array}{l}
\vec{X}_{1} \\
\vec{X}_{2}
\end{array}
\right) =\left(
\begin{array}{ll}
\left( \sqrt{\frac{m}{M_{1}}}\right) \cos \phi & \left( \sqrt{\frac{m}{M_{1}}%
}\right) \sin \phi \\
-\left( \sqrt{\frac{m}{M_{2}}}\right) \sin \phi & \left( \sqrt{\frac{m}{M_{2}%
}}\right) \cos \phi
\end{array}
\right) \left(
\begin{array}{l}
\vec{Y}_{1} \\
\vec{Y}_{2}
\end{array}
\right) ,\,\,\vec{X}_{3}\equiv \vec{Y}_{3},  \label{t2}
\end{equation}

\noindent where $m$ is an arbitrary parameter with dimensions of mass, and $%
\phi $ is the rotation angle which are going to be fixed in order to
disentangle the system.

After using these transformations, one achieves the following Lagrangian for
the last two variables

\begin{eqnarray}
{\cal L} &=&\frac{M}{2}\left( \frac{\,d\vec{Y}_{3}}{dt}\right)
^{2}+\sum_{j=1}^{2}\frac{m}{2}\left( \frac{\,d\vec{Y}_{j}}{dt}\right) ^{2}-%
\frac{1}{2}\left[ \alpha \,\vec{Y}_{1}^{2}+\beta \,\vec{Y}_{2}^{2}+\gamma \,%
\vec{Y}_{1}\cdot \vec{Y}_{2}+\right.  \nonumber \\
&& \\
&&\left. +\,\vec{F}_{1}\left( t\right) \cdot \vec{Y}_{1}+\vec{F}_{2}\left(
t\right) \cdot \vec{Y}_{2}\right] ,  \nonumber
\end{eqnarray}

\noindent with the transformed constant couplings being defined as
\begin{eqnarray}
\alpha &\equiv &m\,\omega _{1}^{2}\,\cos \phi ^{2}+m\,\omega _{2}^{2}\,\sin
\phi ^{2}-\frac{\lambda \,m}{\sqrt{M_{1}M_{2}}}\,\sin \left( 2\phi \right) ;
\nonumber \\
\beta &\equiv &m\,\omega _{1}^{2}\,\sin \phi ^{2}+m\,\omega _{2}^{2}\,\cos
\phi ^{2}+\frac{\lambda \,m}{\sqrt{M_{1}M_{2}}}\,\sin \left( 2\phi \right) ;
\\
\gamma &\equiv &m\,\left( \omega _{1}^{2}\,-\,\omega _{2}^{2}\right) \,\sin
\left( 2\phi \right) +\frac{2\,\lambda \,m}{\sqrt{M_{1}M_{2}}}\,\cos \left(
2\phi \right) ;  \nonumber
\end{eqnarray}

\noindent and
\begin{eqnarray}
\vec{F}_{1} &\equiv &\sqrt{\frac{m}{M_{1}}}\,\vec{f}_{1}\,\cos \phi -\sqrt{%
\frac{m}{M_{2}}}\,\vec{f}_{2}\,\sin \phi ;  \nonumber \\
&& \\
\vec{F}_{2} &\equiv &\sqrt{\frac{m}{M_{1}}}\,\vec{f}_{1}\,\sin \phi +\sqrt{%
\frac{m}{M_{2}}}\,\vec{f}_{2}\,\cos \phi .  \nonumber
\end{eqnarray}

\noindent Furthermore, due to the above transformation we get a change of
the path-integral measure given by a cubic power of the Jacobian of the
transformation ($J^{3}=\left( \frac{\sqrt{M_{1}M_{2}}}{m}\right) ^{3}$). Now
we are in conditions to eliminate the crossed term in these last
coordinates, and this is done by imposing that the angle $\phi $ should
obeys:
\begin{equation}
\tan \left( \phi \right) =\frac{2\,\lambda }{\sqrt{M_{1}M_{2}}}\left( \omega
_{2}^{2}-\omega _{1}^{2}\right) .
\end{equation}

\noindent Solving the above equation, one can easily to note that two
solutions appear, but they just interchange the role of the new vectors $%
\vec{Y}_{1}$ and $\vec{Y}_{2}$ in the Lagrangian, which lead us to conclude
that both conduce to the same physical consequences, in such a way that we
only need to work with one of them. We will use the following solution
\begin{equation}
\cos \phi = \frac{1}{2}\left( 1+\frac{\sqrt{M_{1}M_{2}\left( \omega
_{2}^{2}-\omega _{1}^{2}\right) ^{2}}}{\sqrt{4\,\lambda
^{2}+M_{1}M_{2}\left( \omega _{2}^{2}-\omega _{1}^{2}\right) ^{2}}}\right) .
\end{equation}

Using the above solution, the Lagrangian will be finally set decoupled and
given by

\begin{equation}
{\cal L}=\frac{m}{2}\sum_{j=1}^{2}\left[ \left( \frac{\,d\vec{Y}_{j}}{dt}%
\right) ^{2}-\Omega _{j}^{2}\,\vec{Y}_{j}^{2}+\,\frac{2}{m}\vec{F}_{j}\left(
t\right) \cdot \vec{Y}_{j}\right] ,
\end{equation}

\noindent where the final decoupled frequencies can be written as
\begin{eqnarray}
\Omega _{1}^{2} &\equiv &\frac{1}{2}\left\{ \omega _{1}^{2}+\omega
_{2}^{2}-\left[ \left( \omega _{2}^{2}-\omega _{1}^{2}\right) ^{2}+\frac{%
4\,\lambda }{M_{1}M_{2}}\right] ^{\frac{1}{2}}\right\} ,  \nonumber \\
&& \\
\Omega _{2}^{2} &\equiv &\frac{1}{2}\left\{ \omega _{1}^{2}+\omega
_{2}^{2}+\left[ \left( \omega _{2}^{2}-\omega _{1}^{2}\right) ^{2}+\frac{%
4\,\lambda }{M_{1}M_{2}}\right] ^{\frac{1}{2}}\right\} ,  \nonumber
\end{eqnarray}

\noindent and the forces
\begin{eqnarray}
\vec{F}_{1}\left( t\right) &\equiv &\sqrt{\frac{m}{2\,M_{1}}\left(
1+R\right) }\,\vec{f}_{1}\left( t\right) \,-\sqrt{\frac{m}{2\,M_{2}}\left(
1-R\right) }\,\vec{f}_{2}\left( t\right) \,,  \nonumber \\
&& \\
\vec{F}_{2}\left( t\right) &\equiv &\sqrt{\frac{m}{2\,M_{1}}\left(
1-R\right) }\,\vec{f}_{1}\left( t\right) \,+\sqrt{\frac{m}{2\,M_{2}}\left(
1+R\right) }\,\vec{f}_{2}\left( t\right) \,,  \nonumber
\end{eqnarray}

\noindent where we defined that
\begin{equation}
R\equiv \frac{\sqrt{M_{1}M_{2}\left( \omega _{2}^{2}-\omega _{1}^{2}\right)
^{2}}}{\sqrt{4\,\lambda ^{2}+M_{1}M_{2}\left( \omega _{2}^{2}-\omega
_{1}^{2}\right) ^{2}}}.
\end{equation}
Finally we are left with the task of solving the corresponding Feynman
propagator for a system of uncoupled forced harmonic oscillators. For this
we can use the well known solutions of this system \cite{feynman}.

Remembering the fact that the final Lagrangian is a direct sum of
independent ones, the corresponding propagator must be just the product of
three independent propagators in terms of the final variables, so we get
\begin{eqnarray}
K\left( \vec{Y}_{1}^{,,},\vec{Y}_{1}^{,};\vec{Y}_{2}^{,,},\vec{Y}_{2}^{,};%
\vec{Y}_{3}^{,,},\vec{Y}_{3}^{,};\tau \right) &=&\,\left( a\,b\,\frac{\sqrt{%
M_{1}M_{2}}}{m}\right) ^{3}K_{1}\left( \vec{Y}_{1}^{,,},\vec{Y}_{1}^{,};\tau
\right) \times \nonumber \\
&&\times \,K_{2}\left( \vec{Y}_{2}^{,,},\vec{Y}_{2}^{,};\tau
\right) \,K_{3}\left( \vec{Y}_{3}^{,,},\vec{Y}_{3}^{,};\tau
\right) ,
\end{eqnarray}

\noindent where the prime and the double prime denotes that a function is
being evaluated at the initial or at the final instant of the time interval
respectively. Besides $\vec{\tau}\equiv t_{b}-t_{a}$ is the time interval
between these instants. The first propagator $K_{3}$ is that of a free
three-dimensional particle with the total mass of the system of three
particles, and can be easily obtained \cite{feynman}, as
\begin{equation}
K_{3}\left( \vec{Y}_{3}^{,,},\vec{Y}_{3}^{,};\tau \right) =\left( \frac{M}{%
2\,\pi \,i\,\tau }\right) ^{\frac{3}{2}}\exp \left\{ \frac{i\,M}{2\,\hbar
\,\tau }\left( \vec{X}_{3}^{,,}-\vec{X}_{3}^{,}\right) ^{2}\right\} ,
\end{equation}
once $\vec{Y}_{3}=\vec{X}_{3}$. The other two have the same form of a
three-dimensional driven harmonic oscillator \cite{feynman}:
\begin{eqnarray}
K_{j}\left( \vec{Y}_{j}^{,,},\vec{Y}_{j}^{,};\tau \right) &=&\left[ \,\frac{%
m\,\Omega _{j}}{2\,\pi \,\hbar \,\,i\,\,\sin \left( \Omega _{j}\,\tau
\right) }\right] ^{\frac{3}{2}}\exp \left\{ \frac{i\,\,m\,\Omega _{j}}{%
2\,\,\hbar \,\,\sin \left( \Omega _{j}\,\tau \right) }\left[ \left( \left(
\vec{Y}_{j}^{,,}\right) ^{2}+\right. \right. \right.  \nonumber \\
&& \\
&&\left. \left. \left. +\left( \vec{Y}_{j}^{,}\right) ^{2}\right) \cos
\left( \Omega _{j}\,\tau \right) -2\,\vec{Y}_{j}^{,,}\cdot \vec{Y}%
_{j}^{,}-G_{j}\left( \tau \right) \right] \right\}  \nonumber
\end{eqnarray}

\noindent where
\begin{eqnarray}
G_j\left( \tau \right) &\equiv &\left( \frac{2\,}{m\,\Omega_j }\right)
\left\{ \int_{t_{b}}^{t_{a}}dt\,\left[ \left( \vec{Y}_{j}^{,,}\cdot \vec{F}%
_{j}\left( t\right) \right) \,\sin \left[ \Omega_j \left( \,t-t_{a}\right)
\right] +\right. \right.  \nonumber \\
&&\left. \left. +\left( \vec{Y}_{j}^{,}\cdot \,\vec{F}_{j}\left( t\right)
\right) \,\sin \left[ \Omega_j \left( \,t_{b}-t\right) \right] \right]
\right\} + \\
&&-\left( \frac{2\,}{m^{2}\,\Omega_j ^{2}}\right)
\int_{t_{a}}^{t_{b}}dt^{\prime }\int_{t_{a}}^{t}dt\,\left( \vec{F}_{j}\left(
t\right) \, \cdot \, \vec{F}_{j}\left( t^{\prime }\right) \right) \,\times
\nonumber \\
&&\times \sin \left[ \Omega_j \left( \,t_{b}-t\right) \right] \sin \left[
\Omega_j \,\left( t^{\prime }-t_{a}\right) \right] ,  \nonumber
\end{eqnarray}

\noindent and $j=1,\,2$. Now, by using the above expression, one can finally
write the final expression of the quantum propagator of the three-body
system under examination, simply multiplying the three resulting
propagators, and further taking back the original physical variables,
through
\begin{equation}
\left(
\begin{array}{l}
\vec{Y}_{1} \\
\vec{Y}_{2}
\end{array}
\right) =\left(
\begin{array}{ll}
\left( \sqrt{\frac{M_{1}}{m}}\right) \cos \phi & -\left( \sqrt{\frac{M_{2}}{m%
}}\right) \sin \phi \\
\left( \sqrt{\frac{M_{1}}{m}}\right) \sin \phi & \left( \sqrt{\frac{M_{2}}{m}%
}\right) \cos \phi
\end{array}
\right) \left(
\begin{array}{l}
\vec{X}_{1} \\
\vec{X}_{2}
\end{array}
\right) ,
\end{equation}

\noindent and, also using the initial transformation (\ref{t1}). Here,
however, we avoid to write the final expression for the sake of conciseness
and, because at this point it is only a simple task of substituting and
multiplying the decoupled propagators.

From the above expressions we can extract the corresponding wave functions
and energies. For this we remember that the propagator can be obtained from
the following spectral summation
\begin{equation}
K\left( z^{^{\prime \prime }},z^{\prime };\tau ,0\right) =\sum_{n=0}^{\infty
}\psi _{n}^{*}\left( z^{\prime },t^{\prime }\right) \,\psi _{n}\left(
z^{^{\prime \prime }},t^{^{\prime \prime }}\right) .  \label{decomposition}
\end{equation}

\noindent On the other hand, we can use the Mehler's formula \cite{erdelyi},

\begin{equation}
\frac{\exp \left[ -\left( a^{2}+b^{2}-2\,a\,b\,c\right) /\left(
1-c^{2}\right) \right] }{\sqrt{\left( 1-c^{2}\right) }}=\exp \left[ -\left(
x^{2}+b^{2}\right) \right] \sum_{n=0}^{\infty }\frac{c^{n}}{n!}H_{n}\left(
a\right) H_{n}\left( b\right) ,  \label{mehler}
\end{equation}

\noindent in order to \cite{dutratransf} recover the corresponding wave
functions. After length but straightforward calculations one can finally
obtain a wave function which consists of a three-dimensional free particle
with continuous energy, and three three-dimensional driven harmonic
oscillators, with discrete eigenenergies. Once more, in order to be concise,
we will write below only the quantized part of the wave function,
\begin{equation}
\Psi =\Pi _{a=1}^{3}\psi _{n_{1a}n_{2a}}\left( a_{1},a_{2}\right) ,
\end{equation}

\noindent with
\begin{eqnarray}
\psi _{n_{1a}n_{2a}}\left( a_{1},a_{2}\right)  &=&\frac{1}{2^{-\left(
n_{1a}+n_{2a}\right) }\left( n_{1a}!\,n_{2a}!\right) ^{\frac{1}{2}}}\left(
\frac{M_{1}M_{2}\Omega _{1}\Omega _{2}}{\pi ^{2}\hbar ^{2}}\right) ^{\frac{1%
}{4}}\exp \left( -i\,E_{n_{1a}n_{2a}}\,t\right)   \nonumber \\
&&\times \,\exp \left\{ \left( \frac{1}{2\,\hbar }\right) \left[ -\Omega
_{1}\left( \sqrt{M_{1}}\,C\,a_{1}-\sqrt{M_{2}}\,S\,\,a_{2}+\eta _{1a}\right)
^{2}\right] \right.   \nonumber \\
&&\left. -\Omega _{2}\,\left( \sqrt{M_{1}}\,S\,a_{1}-\sqrt{M_{2}}%
\,C\,a_{2}+\eta _{2a}\right) ^{2}+\right.   \nonumber \\
&&\left. -i\,\dot{\eta}_{1a}\left[ \eta _{1a}-2\left( \sqrt{M_{1}}\,C\,a_{1}-%
\sqrt{M_{2}}\,S\,a_{2}\right) \right] +\right.   \nonumber \\
&&\left. -i\,\dot{\eta}_{2a}\left[ \eta _{2a}-2\left( \sqrt{M_{1}}%
\,S\,\,a_{1}-\sqrt{M_{2}}\,C\,a_{2}\right) \right] \right\}   \nonumber \\
&&\times \exp \left\{ -\frac{i}{2\hbar }\int^{t}d\lambda \left[ \left( \frac{%
\eta _{1a}}{\sqrt{M_{1}}}\left( F_{1a}\,C-f_{2}\,S\right) \right) +\right.
\right.   \nonumber \\
&&\left. \left. \left( \frac{\eta _{2a}}{\sqrt{M_{2}}}\left(
F_{2a}\,S+f_{2}\,C\right) \right) \right] \right\}   \nonumber \\
&&\times H_{n_{1a}}\left[ \sqrt{\frac{\Omega _{1}}{\hbar }}\,\left( \sqrt{%
M_{1}}\,C\,a_{1}-\sqrt{M_{2}}\,S\,a_{2}\right) \right] \,  \nonumber \\
&&\times H_{n_{2a}}\left[ \sqrt{\frac{\Omega _{2}}{\hbar }}\,\left( \sqrt{%
M_{1}}\,S\,a_{1}+\sqrt{M_{2}}\,C\,a_{2}\right) \right] ,
\end{eqnarray}

\noindent where $a=1,2,3=x,y,z$, $S\equiv \sin \left( \phi \right) $, $%
C\equiv \cos \left( \phi \right) $, $H_{n_{i\, a}}(\cdot )$ is the $n_{i\, a
}$ th Hermite polynomial, and
\begin{eqnarray}
\eta _{ia}\left( t\right) &\equiv &\left( \frac{1}{\sqrt{m}\,\sin \left(
\Omega _{i}\,t\right) }\right) \left\{ \int_{t_{a}}^{t}d\xi \,F_{ia}\left(
\xi \right) \,\sin \left( \Omega _{i}\left( \xi -t_{a}\right) \right) \sin
\left( \Omega _{i}\left( t_{b+}-t\right) \right) \right. \\
&&\left. +\int_{t}^{t_{b}}d\xi \,F_{ia}\left( \xi \right) \,\sin \left(
\Omega _{i}\left( \xi -t_{a}\right) \right) \sin \left( \Omega _{i}\left(
t_{b}-t\right) \right) \right\} .  \nonumber
\end{eqnarray}

\noindent Particularly in the case of non-driving forces ($\vec{g}_{1}=\vec{g%
}_{2}=$ $\vec{g}_{3}=0$), the eigen-energies are given by
\begin{equation}
E=\sum_{a=1}^{3}\,E_{n_{1a}n_{2a}}=\sum_{a=1}^{3}\left[ \left( n_{1a}+\frac{1%
}{2}\right) \hbar \,\Omega _{1}+\left( n_{2a}+\frac{1}{2}\right) \hbar
\,\Omega _{2}\right] ,
\end{equation}

\noindent from which we can observe that, beyond the usual degeneracy which
happens when $\sum_{a=1}^{3}n_{1a}=N_{1}$, or $\sum_{a=1}^{3}n_{2a}=N_{2}$,
with $N_1$ and $N_2$ being integer numbers. In the case where the final
frequencies have a rational relation, further degeneracies will appear. As
the transformed frequencies are functions of the original ones and of the
couplings, one concludes that, when certain relations between the parameters
hold, the system becomes more degenerate, signalizing the appearance of
hidden symmetries.

\section{A general exactly solvable quadratic interactions for three-body
with time-dependent couplings}

In this section, we extend our calculation in order to include the case of
time-dependent couplings in the three-body system. In this case the number
of works in this matter is even more scarce and, up to our knowledge,
restrict to the one-dimensional case \cite{khandekar}, \cite{govaerts}. In
this new situation, as a consequence of the fact that if we try to do the
same above transformations, additional terms would appear rendering the
system unsolvable if the masses were time-dependent and we must keep the
masses constant. Furthermore, we should restrict ourselves to treat the case
of isotropic frequencies, and perform some new transformations because the
decoupled system is still time-dependent \cite{dutraPLA90}. In view of these
arguments, we deal with a system characterized by the Lagrangian density (%
\ref{l1}) with arbitrary constant masses, perform the transformations (\ref
{t1}) getting a Lagrangian density with the same form of (\ref{l2}) but now,
one has time-dependent frequencies and couplings. However, in order to
guarantee the exact solvability we must impose the following set of
constraints
\begin{equation}
M_{1}\equiv \frac{m_{1}m_{2}}{a^{2}m_{12}}\,=\,\,M_{2}\equiv \frac{%
m_{3}\,m_{12}}{b^{2}\,M}\equiv \mu ,\,\omega _{1}\left( t\right)
^{2}=\,\omega _{2}\left( t\right) ^{2},
\end{equation}

\noindent which leads us to fix one of the arbitrary constants through the
relation
\begin{equation}
\left( \frac{a}{b}\right) \equiv \sqrt{\frac{m_{1}m_{2}M}{m_{3}m_{12}^{2}}}.
\end{equation}

\noindent On the other hand, the constraint among the frequencies implies
into a restriction over the time-dependency of one of the coupling
parameters, given by
\begin{eqnarray}
\omega ^{2}\left( t\right) &\equiv &\omega _{1}^{2}\equiv \frac{m_{12}}{%
m_{1}m_{2}}\left\{ K_{21}+\frac{1}{m_{12}^{2}}\left(
m_{2}^{2}K_{31}+m_{1}^{2}K_{32}\right) +\right.  \nonumber \\
&&\left. +\frac{2}{m_{12}}\left[ \sigma _{1}m_{2}-\sigma _{2}m_{1}-\sigma
_{3}\left( \frac{m_{1}m_{2}}{m_{12}}\right) \right] \right\} = \\
&=&\omega _{2}^{2}\equiv \frac{M}{m_{3}m_{12}}\left( K_{31}+K_{32}+2\,\sigma
_{3}\right) .  \nonumber
\end{eqnarray}

\noindent Leading to the following Lagrangian
\begin{eqnarray}
{\cal L} &=&\frac{M}{2}\left( \frac{\,d\vec{X}_{3}}{dt}\right)
^{2}+\sum_{j=1}^{2}\frac{\mu }{2}\left( \frac{\,d\vec{X}_{j}}{dt}\right)
^{2}-\frac{1}{2}\left[ \mu \,\omega \left( t\right) ^{2}\left( \,\bar{X}%
_{1}^{2}+\vec{X}_{2}^{2}\right) +\right.  \nonumber \\
&& \\
&&\left. +\,\lambda \left( t\right) \,\,\,\vec{X}_{1}\cdot \vec{X}_{2}+\,%
\vec{f}_{1}\left( t\right) \cdot \vec{X}_{1}+\vec{f}_{2}\left( t\right)
\cdot \vec{X}_{2}\right] ,  \nonumber
\end{eqnarray}

Now, we can decouple the coordinates $\vec{X}_{1}$ and $\vec{X}_{2}$, by
performing a $\pi /4$ rotation around the third vector like
\begin{equation}
\left(
\begin{array}{l}
\vec{X}_{1} \\
\vec{X}_{2}
\end{array}
\right) =\frac{1}{\sqrt{2}}\left(
\begin{array}{ll}
1 & 1 \\
-1 & 1
\end{array}
\right) \left(
\begin{array}{l}
\vec{x}_{1} \\
\vec{x}_{2}
\end{array}
\right) ,
\end{equation}

\noindent and this leads us to the following Lagrangian
\begin{eqnarray}
{\cal L} &=&\frac{M}{2}\left( \frac{\,d\vec{X}_{3}}{dt}\right)
^{2}+\sum_{j=1}^{2}\frac{\mu }{2}\left( \frac{\,d\vec{x}_{j}}{dt}\right)
^{2}-\frac{\mu }{2}\left[ \Omega _{1}\left( t\right) ^{2}\,\vec{x}%
_{1}^{2}+\Omega _{2}\left( t\right) ^{2}\vec{x}_{2}^{2}\right.  \nonumber \\
&& \\
&&\left. +\vec{\theta}_{1}\left( t\right) \cdot \vec{x}_{1}+\vec{\theta}%
_{2}\left( t\right) \cdot \vec{x}_{2}\right] ,  \nonumber
\end{eqnarray}

\noindent where the frequencies and linear force-type couplings of the
decoupled three-dimensional oscillators are respectively given by
\begin{equation}
\Omega _{i}\left( t\right) ^{2}\equiv \omega \left( t\right) ^{2}\pm \frac{%
\lambda \left( t\right) }{\mu };\,\,\vec{\theta}_{i}\left( t\right) \equiv
\frac{1}{\sqrt{2}}\left( \,\vec{f}_{1}\left( t\right) \pm \,\vec{f}%
_{2}\left( t\right) \right) ;\,i=1,2.
\end{equation}

Furthermore, the path integral measure is invariant under this
transformation. As a consequence, we are left to solve the problem of a free
particle (coordinate $\vec{X}_{3}$), and two forced oscillators with
time-dependent frequencies and forces \cite{dutraPLA90}. Once the case of
the oscillators is more involved, we review briefly a way to get the
propagator for them. First of all, we perform a translation like
\begin{equation}
\vec{y}_{i}\,\equiv \,\vec{x}_{i}\,+\vec{\eta}_{i}\,,
\end{equation}

\noindent and then impose the elimination of the linear terms, which make it
necessary to restrict the arbitrary functions $\vec{\eta}_{i}$ through the
equation
\begin{equation}
\frac{d^{2}\vec{\eta}_{i}}{dt^{2}}+\Omega _{i}\left( t\right) ^{2}\,\vec{\eta%
}_{i}=-\frac{\vec{\theta}_{i}}{\mu },
\end{equation}

\noindent and this allow us to write the Lagrangians as
\begin{equation}
{\cal L}_{j}=\frac{\mu }{2}\left( \frac{\,d\vec{y}_{j}}{dt}\right) ^{2}-%
\frac{\mu }{2}\,\,\Omega _{j}\left( t\right) ^{2}\,\vec{y}_{j}^{2}\,+\frac{%
d\,\vec{F}_{j}}{dt}\,;\,\,j=1,2
\end{equation}

\noindent with
\begin{equation}
\frac{d\,\vec{F}_{j}}{dt}\equiv \frac{1}{2}\mu \,\frac{d\vec{\eta}_{j}}{dt}%
\,\cdot \left( \vec{\eta}_{j}-2\,\,\vec{y}_{j}\right) -\frac{1}{2}\int^{t}%
\vec{\theta}_{j}\left( \xi \right) \,\cdot \vec{\eta}_{j}\left( \xi \right)
\,d\xi ,
\end{equation}

From the above, we conclude that the translation used has been able to
reduce the problem to that of a harmonic oscillator with time-dependent
frequency. At this point we can, for instance, map the problem into that of
a free particle, what can be done by using a set of coordinate
transformations, including a time substitution, introduced by Cheng many
years ago \cite{cheng2}, \cite{Cheng},
\begin{equation}
\vec{z}_{i}\equiv \vec{y}_{i}\,\,\dot{\alpha}_{i}^{1/2}\,\sec \left[ \mu
_{i}\left( t\right) \right] ,\,\,u_{i}\equiv \tan \left[ \alpha _{i}\left(
t\right) \right] ,
\end{equation}

\noindent where $\alpha _{i}\left( t\right) $ and some auxiliary variables $%
s_{i}\left( t\right) $ should obey the equations
\begin{equation}
\ddot{s}_{i}+\Omega _{i}\left( t\right) ^{2}\,s_{i}=\frac{1}{s_{i}^{3}}%
,\,\,s_{i}^{2}\,\dot{\alpha}_{i}=1.
\end{equation}

After these transformations, we ends with the following Lagrangians
\begin{equation}
{\cal L}_{j}=\frac{\mu }{2}\left( \frac{d\vec{z}_{j}}{du_{j}}\right) ^{2}+%
\frac{d\,\vec{F}_{j}\left( \vec{y}_{i},\vec{\eta}_{i},t\right) }{dt}-\,\frac{%
dG_{j}\left( s_{i},t\right) }{dt},\,
\end{equation}

\noindent with
\begin{equation}
G_{j}\left( s_{i},t\right) \equiv \frac{1}{2}\mu \,\vec{y}_{j}^{2}\left\{
\sin \left[ 2\,\mu _{j}\left( t\right) \right] -2\frac{\ddot{s}_{j}}{s_{j}\,%
\dot{u}_{j}}\right\} .
\end{equation}

Looking to the above equations, one can conclude that we have really mapped
the original problem into that of free particles. Now, using the solution of
the free particle oscillator \cite{feynman} and the Van Vleck-Pauli formula
\cite{vanvleck}, it can be shown after straightforward calculations that the
$i_{th}$ propagator becomes
\begin{eqnarray}
K_{ia}\left( y_{ia}^{^{\prime \prime }},y_{ia}^{\prime };\tau ,0\right)
&=&\left( \frac{\sqrt{\dot{\alpha}_{ia}^{\prime }\alpha _{ia}^{"}}}{2\,\pi
\,i\,\sin \delta _{ia}}\right) ^{\frac{1}{2}}\exp \left[ \frac{i\,\mu }{%
2\,\hbar }\,\left( y_{ia}^{2}\frac{\dot{s}_{ia}}{s_{ia}}+\dot{\eta}%
_{ia}\left( \eta _{ia}-2\,y_{ia}\right) \right) _{0}^{\tau }\right] \times
\nonumber \\
&&\times \,\exp \left\{ \frac{i\,\mu }{2\,\hbar \,\sin \delta _{ia}}\left[
\left( \,\alpha _{ia}^{"}\,\left( y_{ia}"\right) ^{2}+\,\dot{\alpha}%
_{ia}^{\prime }\left( \,y_{ia}^{\prime }\right) ^{2}\right) \cos \delta
_{ia}+\right. \right.  \nonumber \\
&&\left. \left. -2\,\,\,\sqrt{\dot{\alpha}_{ia}^{\prime }\alpha _{ia}^{"}}%
\,\,y_{ia}^{\prime }\,\,y_{ia}"\right] \right\} \,\times \,  \nonumber \\
&&\times \exp \left( -\frac{i}{2\,\hbar }\int_{0}^{\tau }\,\theta
_{ia}\left( \xi \right) \eta _{ia}\left( \xi \right) \,d\xi \right) ,
\end{eqnarray}

\noindent where it was define for the sake of compactification of the
expressions that, for any function $g\left( t\right) $, one have that $%
g\left( \tau \right) \equiv g"$ and $g\left( 0\right) \equiv g^{\prime }$.
Besides, in the above expression $\delta _{ia}\equiv \alpha _{ia}^{"}-\dot{%
\alpha}_{ia}^{\prime }$.
\begin{eqnarray}
K\left( \vec{Y}_{1}^{,,},\vec{Y}_{1}^{,};\vec{Y}_{2}^{,,},\vec{Y}_{2}^{,};%
\vec{Y}_{3}^{,,},\vec{Y}_{3}^{,};\tau \right) &=&\,\left( b^{2}\sqrt{\frac{%
m_{1}m_{2}M_{1}M_{2}}{m_{3}m_{12}^{2}M}}\,\right) ^{3}\left( \frac{M}{2\,\pi
\,i\,\tau }\right) ^{\frac{3}{2}}\times  \nonumber \\
&&\times \exp \left\{ \frac{i\,M}{2\,\hbar \,\tau }\left( \vec{X}_{3}^{,,}-%
\vec{X}_{3}^{,}\right) ^{2}\right\} \\
&&\times \Pi _{i=1}^{2}\Pi _{a=1}^{3}\,K_{ia}\left( \vec{y}_{ia}^{,,},\vec{y}%
_{ia}^{,};\tau \right) .  \nonumber
\end{eqnarray}

Let us now recover the wave functions through the use of the decomposition
of the propagator in terms of the wave functions, as it is given from (\ref
{decomposition}), and again using the Mehler's formula (\ref{mehler}), with
the necessary identifications $a_{i}=\sqrt{\frac{\mu \,\dot{\alpha}_{i}^{%
\acute{}}}{\hbar }}\,y_{i}^{\acute{}\acute{}}$, $b_{i}=\sqrt{\frac{\mu \,%
\dot{\alpha}_{i}^{\acute{}\acute{}}}{\hbar }}\,y_{i}^{\acute{}}$ and $%
c_{i}=\exp \left( -i\,\delta _{i}\right) $, we obtain for the discrete part
of the wave functions
\begin{equation}
\Psi \left( \vec{X}_{1},\vec{X}_{2},\vec{X}_{3}\right) =\Pi _{a=1}^{3}\psi
_{n_{1a},n_{2a}}\left( X_{1a},X_{2a}\right) \,\exp \left[ i\,\vec{k}\cdot \,%
\vec{X}_{3}\right] ,
\end{equation}

\noindent with
\begin{eqnarray}
\psi _{n_{1a},n_{2a}}\left( X_{1a},X_{2a}\right) &=&\left[ \frac{1}{%
2^{n_{1a}+n_{2a}}\left( n_{1a}!\,n_{2a}!\right) }\left( \frac{\mu ^{2}\,\dot{%
\alpha}_{1\,}\dot{\alpha}_{2}}{\left( \pi \,\hbar \right) ^{2}}\right)
\right] ^{\frac{1}{2}}  \nonumber \\
&&\times \exp \left\{ \frac{i\,\mu }{4\,\hbar }\left[ \left( X_{1a}-X_{2a}+%
\sqrt{2}\,\eta _{1a}\right) ^{2}\left( \frac{\dot{s}_{1}}{s_{1}}-\dot{\alpha}%
_{1}\right) + \right. \right.  \nonumber \\
&&\left. \left. -2\dot{\eta}_{1a}\left( \eta _{1a}+\sqrt{2}\left(
X_{1a}-X_{2a}\right) \right) \right] \right\}  \nonumber \\
&&\times \exp \left\{ \frac{i\,\mu }{4\,\hbar }\left[ \left( X_{1a}+X_{2a}+%
\sqrt{2}\,\eta _{2a}\right) ^{2}\left( \frac{\dot{s}_{2}}{s_{2}}-\dot{\alpha}%
_{2}\right) \right. \right. +  \nonumber \\
&&\left. \left. -2\dot{\eta}_{2a}\left( \eta _{2a}+\sqrt{2}\left(
X_{1a}+X_{2a}\right) \right) \right] \right\}  \nonumber \\
&&\times H_{n_{a1}}\left( \left[ \frac{\mu \,\dot{\alpha}_{1}}{2\,\hbar }%
\right] ^{\frac{1}{2}}\left( X_{1a}-X_{2a}+\sqrt{2}\,\eta _{1a}\right)
\right) \; \\
&&\times H_{n_{a2}}\left( \left[ \frac{\mu \,\dot{\alpha}_{2}}{2\,\hbar }%
\right] ^{\frac{1}{2}}\left( X_{1a}+X_{2a}+\sqrt{2}\,\eta _{2a}\right)
\right)  \nonumber \\
&&\times \exp \left[ -\frac{i}{2\,\,\hbar }\int^{\tau }\,\left( \theta
_{1a}\left( \xi \right) \eta _{1a}\left( \xi \right) +\theta _{2a}\left( \xi
\right) \eta _{2a}\left( \xi \right) \right) \,d\xi \right]  \nonumber \\
&&\times \exp \left\{ -i\left[ \left( n_{1a}+\frac{1}{2}\right) \alpha
_{1}\right] +\left( n_{2a}+\frac{1}{2}\right) \alpha _{2}\right\} .
\nonumber
\end{eqnarray}

Finally we would like to remark that some other systems which are somewhat
more general than that appeared until now in the literature could be
included in this analysis. This is the case, for instance, of a
three-dimensional generalization of the potential considered in the
interesting work of Chouchaoui \cite{3bodes}, which would be represented by
the following Lagrangian
\begin{eqnarray}
{\cal L} &=&\sum_{j=1}^{3}\frac{m_{j}}{2}\left( \frac{\,d\vec{r}_{j}}{dt}%
\right) ^{2}-\frac{1}{2}\left[ K_{21}\,\left( \vec{r}_{2}-\vec{r}_{1}\right)
^{2}+K_{31}\,\left( \vec{r}_{3}-\vec{r}_{1}\right) ^{2}+K_{32}\,\left( \vec{r%
}_{3}-\vec{r}_{2}\right) ^{2}+\right.   \nonumber \\
&&+\,\sigma _{1}\left( \vec{r}_{2}-\vec{r}_{1}\right) \cdot \left( \vec{r}%
_{3}-\vec{r}_{1}\right) +\sigma _{2}\left( \vec{r}_{2}-\vec{r}_{1}\right)
\cdot \left( \vec{r}_{3}-\vec{r}_{2}\right) + \\
&&\left. +\sigma _{3}\left( \vec{r}_{3}-\vec{r}_{1}\right) \cdot \left( \vec{%
r}_{3}-\vec{r}_{2}\right) +\frac{g_{1}}{\left( \vec{r}_{2}-\vec{r}%
_{1}\right) ^{2}}\right] ,  \nonumber
\end{eqnarray}

\noindent and with some restrictions over the potential
parameters. Furthermore, we could also include in the list of the
systems which can be decoupled through the use of the
transformations above described, some cases where particular
three-body interactions could added to the general quadratic
coordinate one. More precisely, one could accrescent something
like
\begin{eqnarray}
\Delta V &=& \frac{g_{1}}{\left( \vec{r}_{2}-\vec{r}_{1}\right)} +\frac{g_{2}%
}{\left( m_{12} \vec{r}_{3}- \left( m_1 \vec{r}_{1}+m_2 \vec{r}_2 \right)
\right)^2} +\frac{g_{3}}{\left( m_1 \vec{r}_{1}+ m_2 \vec{r}_{2}+ m_3 \vec{r}%
_{3} \right)^2 } ,  \nonumber
\end{eqnarray}

\noindent to the interacting potential. But, once again,
restrictions over the parameters of the potential should be
imposed in order to keep the exactness of the propagator and wave
functions solutions. Moreover, the case of the general quadratic
spatial interactions, which we considered in this work, has the
advantage of presenting covariant interactions between the three
particles involved. For this reason, we prefer do not treat these
cases here.

\bigskip

\noindent {\bf Acknowledgments:} The author is grateful to CNPq for partial
financial support. This work has been done during a visit to the ICTP, under
the auspices of the Associate Scheme of the Abdus Salam ICTP.

\end{document}